# Analytical Gradient Theory for Strongly Contracted (SC) and Partially Contracted (PC) *N*-Electron Valence State Perturbation Theory (NEVPT2)

Jae Woo Park[*]

*Department of Chemistry, Chungbuk National University (CBNU), Cheongju 28644, Korea*

**Abstract**

An analytical gradient theory for single-state *N*-electron valence state perturbation theory (NEVPT2), using both strongly contracted (SC) and partially contracted (PC) internal contraction schemes, is developed. We demonstrate the utility of the developed algorithm in the optimization of the single-state molecular geometry of acrolein, benzyne, benzene, the retinal chromophore PSB3, the GFP chromophore *p*HBI, and porphine, with the cc-pVTZ basis sets. The SC-NEVPT2 analytical gradients exhibit numerical instability due to the lack of invariance with respect to the rotations among the inactive orbitals. On the other hand, PC-NEVPT2 gives molecular geometries comparable to CASPT2 in any tested cases. We discuss possible future developments that will make the NEVPT2 gradient algorithm a powerful tool for optimizing the molecular geometries and conducting molecular dynamics simulations of correlated systems.

---

[*] E-mail: jaewoopark@cbnu.ac.kr

# 1. INTRODUCTION

One of the most prominent applications of quantum chemistry is the exploration of potential energy surfaces. Geometry optimizations and *ab initio* molecular dynamics are the mostly commonly used strategies, and for larger molecules, for practical reasons, the nuclear gradient should be evaluated analytically, rather than numerically.

The accuracy of such calculations significantly depends on the selection of the quantum chemistry method. The electronic correlations can be used to describe the accuracy of the quantum chemical method. When the wave function of the system is not appropriately described by the single determinant, the static correlation should be included as appropriate. Multireference methods can be used to describe static correlations in which multiple Slater determinants are included in the calculation of the reference wave function. The most accurate multireference method is full configuration interaction (FCI). In practice, FCI theory cannot be used even for small molecules, although recently emerging computational methods, such as density-matrix renormalization group (DMRG)[1-4] and FCI quantum Monte Carlo (FCI-QMC)[5-7] can open a way towards approximate FCI calculations for large systems (up to 100 electrons in 100 orbitals). The FCI wave function is usually embedded in the mean-field, resulting in the complete active space self-consistent field (CASSCF) method.[8,9] The dynamic correlation, which is the correlation outside the active space, should be described to achieve accurate descriptions of the energy and potential energy surfaces.[10]

Multireference perturbation theories provide an efficient and accurate means of recovering the dynamical correlation energy from the mean-field multireference function. For example, the multireference second-order Møller–Plesset theory (MR-MP2) by Hirao,[11] the multireference open-shell perturbation theory (MROPT) by Davidson and coworkers,[12] and the generalized Van

Vleck perturbation theory by Hoffmann and coworkers[13] have been reported. By using the internally contracted basis set for perturbation, the second-order complete active space perturbation theory (CASPT2) theory was formulated.[14,15] There are two CASPT2 theories with different internal contraction schemes: fully internally contracted (implemented in OpenMolcas,[16,17] BAGEL,[18] MOLPRO,[19] and CHEMPS2[20,21]) and partially internally contracted[22] (implemented in MOLPRO[19]). Moreover, the extension of these theories to multistate systems, such as in (X)MCQDPT2[23,24] and (X)MS-CASPT2[24,25], have became routine tools for studying excited states.

$N$-Electron valence state perturbation theory (NEVPT) is one of the most recent developments.[26-28] NEVPT2 theory is now common in quantum chemistry software programs, such as QCMAQUIS[29,30] (and its interface to OpenMolcas,[16,17] based on the original implementation of NEVPT2 by Angeli et al.,[26,27,31]) MOLPRO,[19] ORCA,[32,33] PYSCF,[34] DALTON,[35] and BAGEL.[18] NEVPT uses the Dyall Hamiltonian[36] to define the zeroth-order Hamiltonian for perturbation rather than the one-electron generalized Fock operator used in CASPT2. This method is size-consistent and avoids the "intruder state" problem because of the two-electron nature of the zeroth-order Hamiltonian.[26] Additionally, the subspaces of the perturbation functions are fully separable, which means that iteration is not required for solving the amplitude equations (which is the same as in MP2). Similar to the (X)MCQDPT2 and (X)MS-CASPT2 methods, the quasidegenerate formation of NEVPT2, QD-NEVPT2,[31] is suitable for studying excited states.

For CASPT2, the nuclear gradient theory was first formulated for the partially internally contracted variant,[25,37,38] and then it was expanded to the fully internally contracted variant using the automatic code generation technique.[39-44] Because of these developments, carrying out geometry optimizations and molecular dynamics (MD) simulations using CASPT2 is now routine and has moderate computational costs.[41,45-47] Other uncontracted multireference perturbation methods, such as MCQDPT2 and GVVPT2, have algorithms for nuclear gradient as well.[48-50] On the other hand, there were no analytical gradient methods for SC- and PC-NEVPT2 (although the authors learned that Nishimoto implemented an analytical PC-NEVPT2 gradient derived by direct differentiation just before the submission of this work[51]), and consequently, the utility of NEVPT2 toward potential energy surface exploration is somewhat limited. Due to the aforementioned desirable properties of NEVPT2 theory, the NEVPT2 method would be a suitable multireference quantum chemistry method for dynamics simulations and geometry optimizations.

There are three internal contraction schemes for NEVPT2: strongly contracted (SC), partially contracted (PC), and fully uncontracted (UC).[26] Due to their efficiency, the SC and PC schemes are usually used in chemical applications. Although the number of basis functions (or size of the perturbation subspaces) is different for each contraction scheme, the resulting potential energy surfaces do not significantly depend on the contraction schemes for either CASPT2[39] or NEVPT2,[26,27] except when the local correlation approximation is used.[52]

In this work, we present a theory and an algorithm for computing the analytical nuclear gradient for both SC and PC single-state NEVPT2. This theory is based on a set of techniques and knowledge accumulated through the development of CASPT2 analytical gradient theory, including the use of the Lagrangian (or response functions),[53-56] which leads to a straightforward derivation. However, we did not rely on an automatic implementation scheme, as the formalism for the NEVPT2 gradient was somewhat simpler than that of CASPT2, mainly because of the lack of off-diagonal elements for the zeroth-order Hamiltonian. We utilized established computational techniques, including the density-fitting (DF) algorithm for two-electron integrals and its nuclear gradients, to make the algorithm efficient. Our results show that the SC-NEVPT2 gradients exhibits numerical instability, due to the lack of invariance with respect to the rotations among the inactive orbitals.[28] On the other hand, the PC-NEVPT2 theory is suitable for optimizing the structures of the large-sized systems, with and without degenerate orbitals. Finally, we compare CASPT2 and NEVPT2 optimized geometries of various organic molecular systems and the computation times required by these systems and discuss future prospects.

## 2. THE NEVPT2 THEORY

First, let us briefly review the NEVPT2 theory such that the analytical gradient theory can be easily formulated. We follow the conventions of the original spin-free NEVPT2[27] and relativistic NEVPT2 references[57] for the orbital indices. The indices $i, j, k, \ldots$; $a, b, c, \ldots$; and $r, s, t, \ldots$ are used to label the closed, active, and virtual orbitals, respectively. In NEVPT2 (regardless of the contraction scheme), the Dyall Hamiltonian ($\hat{H}^D$) is used to define the zeroth-order Hamiltonian.

$$\hat{H}^{(0)} = \hat{P}\hat{H}^D\hat{P} + \hat{Q}\hat{H}^D\hat{Q}, \qquad (1)$$

where the operators $\hat{P}$ and $\hat{Q}$ are the projection operators to the reference space and its complementary space (the first-order interacting space). The Dyall Hamiltonian is

$$\hat{H}^D = \hat{H}_i + \hat{H}_v + C, \qquad (2)$$

where a one-electron operator describes the inactive space as follows.

$$\hat{H}_i = \sum_i^{core} f_{ii}\hat{E}_{ii} + \sum_r^{virt} f_{rr}\hat{E}_{rr}, \qquad (3)$$

A two-electron operator, with a form similar to the full Hamiltonian, treats the active space.

$$\hat{H}_v = \sum_{ab}^{\text{act}} h_{ab}^{\text{eff}} \hat{E}_{ab} + \frac{1}{2} \sum_{abcd}^{\text{act}} \left( \hat{E}_{ac} \hat{E}_{bd} - \delta_{bc} \hat{E}_{ad} \right). \tag{4}$$

Here, $h^{\text{eff}}$ includes the mean-field contributions from the core electrons. Physically, the Dyall Hamiltonian is the same as the Hamiltonian used in the CASSCF calculations; the mean-field approximation describes the inactive space, while the full CI calculation is used for the active space.

The full Hamiltonian is divided into the zeroth-order Hamiltonian and the perturbation as follows:

$$\hat{H} = \hat{H}^{(0)} + \hat{V}. \tag{5}$$

More specifically, the perturbation involves the promotions of the electrons to the virtual or active space from the active or closed space (which is not included in the Dyall Hamiltonian). This perturbation is divided via appropriate partitioning as follows.

$$\hat{V} = \sum_{ijrs} \hat{V}_{ijrs}^{(0)} + \sum_{ijr} \hat{V}_{ijr}^{(+1)} + \sum_{rsi} \hat{V}_{rsi}^{(-1)} + \sum_{ij} \hat{V}_{ij}^{(+2)} + \sum_{rs} \hat{V}_{rs}^{(-2)} + \sum_{i} \hat{V}_{i}^{'(+1)} + \sum_{r} \hat{V}_{r}^{'(-1)} + \sum_{ir} \hat{V}_{ir}^{'(0)}. \tag{6}$$

Depending on the form of the perturbation functions (or the "subspaces" spanned by the perturbation function, $S_l^{(k)}$), the SC-NEVPT2 and PC-NEVPT2 theories are formulated as follows.

**Strongly Contracted NEVPT2 (SC-NEVPT2).** The perturbation function for the SC-NEVPT2 theory is

$$|\Psi_l^{(k)}\rangle = \hat{V}_l^{(k)} |\Psi^{(0)}\rangle. \tag{7}$$

Here, the perturbation operator includes molecular integrals as well as excitation operators. Now, the minimum of the Hylleraas functional,

$$E = 2\langle \Psi^{(1)} | \hat{H} | \Psi^{(0)} \rangle + \langle \Psi^{(1)} | \hat{H}^{(0)} - E^{(0)} | \Psi^{(1)} \rangle \tag{8}$$

is the energy. The first-order wave function is

$$|\Psi^{(1)}\rangle = \sum_{l,k} T_l^{(k)} |\Psi_l^{(k)}\rangle, \tag{9}$$

where $T_l^{(k)}$ is the amplitude for the perturbation function $|\Psi_l^{(k)}\rangle$. The amplitude equation, which is obtained by differentiating the Hylleraas functional with respect to the amplitude, is

$$\langle\Psi_l^{(k)}|\hat{H}|\Psi^{(0)}\rangle + \langle\Psi_l^{(k)}|\hat{H}^{(0)} - E^{(0)}|\Psi_l^{(k)}\rangle T_l^{(k)} = 0. \tag{10}$$

We define the norm of the perturbation function as

$$N_l^{(k)} = \langle\Psi_l^{(k)}|\hat{H}|\Psi^{(0)}\rangle \tag{11}$$

for further convenience. The amplitude is then

$$T_l^{(k)} = -\frac{\langle\Psi_l^{(k)}|\hat{H}|\Psi^{(0)}\rangle}{\langle\Psi_l^{(k)}|\hat{H}^{(0)} - E^{(0)}|\Psi_l^{(k)}\rangle} = -\frac{1}{E_l^{(k)}/N_l^{(k)} - E^{(0)}}, \tag{12}$$

where the diagonal element of the energy is defined as

$$E_l^{(k)} = \langle\Psi_l^{(k)}|\hat{H}^{(0)}|\Psi_l^{(k)}\rangle. \tag{13}$$

The second-order energy is given by

$$E^{(2)} = \sum_{l,k} 2T_l^{(k)} N_l^{(k)} + T_l^{(k)2}(E_l^{(k)} - E^{(0)} N_l^{(k)}), \tag{14}$$

and with the amplitude given by Eq. (12), this can be simplified to

$$E^{(2)} = \sum_{l,k} T_l^{(k)} N_l^{(k)} = -\sum_{l,k} \frac{N_l^{(k)}}{E_l^{(k)}/N_l^{(k)} - E^{(0)}}, \tag{15}$$

which is a form that is presented repeatedly in the literature.[27,28]

**Partially Contracted NEVPT2 (PC-NEVPT2).** The perturbation function for the PC-NEVPT2 theory is

$$|\Psi_{l,\mu}^{(k)}\rangle = V_{A,\mu} \hat{E}_{l,A} |\Psi^{(0)}\rangle, \tag{16}$$

where $A$ includes the appropriate active orbital indices for subspace $S_l^{(k)}$. For example, in the $S_{rs}^{(-2)}$ subspace, $A$ involves two active indices, and the excitation operator and transformation matrices are $\hat{E}_{rb}\hat{E}_{sa}$ and $V_{ab,\mu}$, respectively. The transformation matrix **V** transforms the redundant basis set into the orthogonal basis set, which diagonalizes the elements of Dyall's Hamiltonian in each subspace. Here, the excitation operator, $\hat{E}_{l,A}$, does not include the molecular integrals. Again, the minimum of the Hylleraas functional is the energy. The first-order wave function is

$$|\Psi^{(1)}\rangle = \sum_{l,k}\sum_{\mu} T_{l,\mu}^{(k)} |\Psi_{l,\mu}^{(k)}\rangle, \qquad (17)$$

where $T_{l,\mu}^{(k)}$ is the amplitude of the perturbation function $\Psi_{l,\mu}^{(k)}$. The amplitude equation is then

$$\langle \Psi_{l,\mu}^{(k)} | \hat{H} | \Psi^{(0)} \rangle + \sum_{\nu} \langle \Psi_{l,\mu}^{(k)} | \hat{H}^{(0)} - E^{(0)} | \Psi_{l,\nu}^{(k)} \rangle T_{l,\nu}^{(k)} = 0. \qquad (18)$$

The amplitude equation can be simplified if the orthogonal basis set, which diagonalizes Dyall's Hamiltonian, is used. If that is the case, the amplitude is simply

$$T_{l,\mu}^{(k)} = -\frac{\langle \Psi_{l,\mu}^{(k)} | \hat{H} | \Psi^{(0)} \rangle}{\langle \Psi_{l,\mu}^{(k)} | \hat{H}^{(0)} - E^{(0)} | \Psi_{l,\mu}^{(k)} \rangle}, \qquad (19)$$

and the second-order energy is

$$E^{(2)} = \sum_{l,k}\sum_{\mu} 2T_{l,\mu}^{(k)} \langle \Psi_{l,\mu}^{(k)} | \hat{H} | \Psi^{(0)} \rangle + \sum_{l,k}\sum_{\mu}\sum_{\nu} T_{l,\mu}^{(k)} T_{l,\nu}^{(k)} \langle \Psi_{l,\mu}^{(k)} | \hat{H}^{(0)} - E^{(0)} | \Psi_{l,\nu}^{(k)} \rangle. \qquad (20)$$

Again, the last term can be simplified if the orthogonal basis set is used.

## 3. ANALYTICAL GRADIENT THEORY

Next, let us present the analytical gradient theory for both the SC-NEVPT2 and PC-NEVPT2 theories. The NEVPT2 energy can be written as a functional of amplitude **T**, orbital coefficient **C**, and CI contraction coefficient **c**, as shown by the following:

$$E = E(\mathbf{T}, \mathbf{C}, \mathbf{c}). \tag{21}$$

Then, the nuclear gradient is

$$\frac{d E}{d \mathbf{X}} = \frac{\partial E}{\partial \mathbf{X}} + \frac{\partial E}{\partial \mathbf{T}} \frac{\partial \mathbf{T}}{\partial \mathbf{X}} + \frac{\partial E}{\partial \mathbf{C}} \frac{\partial \mathbf{C}}{\partial \mathbf{X}} + \frac{\partial E}{\partial \mathbf{c}} \frac{\partial \mathbf{c}}{\partial \mathbf{X}}, \tag{22}$$

where **X** is a vector of the nuclear positions. The Hylleraas functional is stationary with respect to the perturbative amplitudes, and as a result, the second term on the right-hand side vanishes. Instead of calculating the derivatives of the coefficients **C** and **c** with respect to the nuclear positions, the Lagrangian formalism[53-56] is used. Simply, the Lagrangian formalism treats all the convergence conditions as constraints, and the corresponding multipliers are evaluated. The NEVPT2 Lagrangian,

$$\begin{aligned}L = E &+ \frac{1}{2} \mathrm{tr}\left[\mathbf{Z}^\dagger (\mathbf{A} - \mathbf{A}^\dagger)\right] - \frac{1}{2} \mathrm{tr}\left[\mathbf{X}(\mathbf{C}^\dagger \mathbf{S}\mathbf{C} - \mathbf{1})\right] \\ &+ \sum_M \mathbf{z}_M^\dagger (\mathbf{H} - E)\mathbf{c}_M - \frac{1}{2} \sum_M x_M \left(\mathbf{c}_M^\dagger \mathbf{c}_M - 1\right) \\ &+ \sum_{i \neq j} z_{ij}^c f_{ij} + \sum_{r \neq s} z_{rs}^c f_{rs}, \end{aligned} \tag{23}$$

is stationary with respect to **C** and **c** when suitable values of **Z** and **z** are used. This Lagrangian takes the same form as that for CASPT2 with the imaginary level shift[44] because of the pseudocanonical conditions. Here, the terms in the first and second lines account for the CASSCF convergence condition, and the last line is due to the pseudocanonical condition for the orbitals ensuring that the Fock operator is diagonal in the core and virtual spaces. The pseudocanonical condition also accounts for the frozen core approximation.[37,44] Then, the so-called Z-vector equations,

$$\frac{\partial L}{\partial \kappa_{xy}} = 0, \tag{24}$$

$$\frac{\partial L}{\partial c_I} = 0, \qquad (25)$$

is solved to obtain the Lagrangian multipliers in the same way as for the SA-CASSCF or CASPT2 gradients.[37]

The source terms for these Z-vector equations,

$$Y_{xy} = \frac{\partial E}{\partial \kappa_{xy}}, \qquad (26)$$

$$y_I = \frac{\partial E}{\partial c_I}, \qquad (27)$$

can be evaluated by taking advantage of the fact that we have to evaluate the derivatives only to the first order. We can write the orbital gradient as

$$\begin{aligned} Y_{xy} = \Big[ & \mathbf{h}\mathbf{d}^{(0)} + \mathbf{h}^{\text{eff}}\mathbf{d}^{\text{eff}} + \mathbf{h}^{\text{eff}'}\mathbf{d}^{\text{eff}'} + \overline{\mathbf{h}}^{\text{eff}}\overline{\mathbf{d}}^{\text{eff}} + \mathbf{f}^h\mathbf{d}^h + \mathbf{f}\mathbf{d}^{\text{Fock}} \\ & + \mathbf{g}\Big(\mathbf{d}^{\text{eff}} + \mathbf{d}^{\text{eff}'} + \overline{\mathbf{d}}^{\text{eff}} + \mathbf{d}^h\Big)\mathbf{d}^{(0)}_{\text{core}} \\ & + \mathbf{g}(\mathbf{d}^{\text{Fock}})\mathbf{d}^{(0)} + \mathbf{g}(\mathbf{d}^h)\mathbf{d}^c_{\text{act}} + \mathbf{g}'\Big(\mathbf{d}^{\text{eff}'} + \frac{1}{2}\overline{\mathbf{d}}^{\text{eff}}\Big)\mathbf{d}^c_{\text{act}} + \sum_{kl}\mathbf{D}^{kl}\mathbf{K}^{lk} \Big]_{xy}, \end{aligned} \qquad (28)$$

where

$$[\mathbf{g}(\mathbf{d})]_{xy} = \sum_{zw}\Big[-\frac{1}{2}(xw|zy) + (xy|zw)\Big]d_{zw}, \qquad (29)$$

$$[\mathbf{g}'(\mathbf{d})]_{xy} = \sum_{zw}\Big[-\frac{1}{2}(xw|zy)\Big]d_{zw}, \qquad (30)$$

$\mathbf{d}^{(0)}$ is the zeroth-order one-particle density matrix, $\mathbf{d}^{(0)}_{\text{core}}$ and $\mathbf{d}^c_{\text{act}}$ are the density matrices with diagonal elements of 2 for the core and active orbitals, respectively, and $\mathbf{d}$ and $\mathbf{D}^{kl}$ are density-like terms ("pseudodensity") for evaluating the source term. These pseudodensities can be computed by simply collecting all the quantities that are multiplied to the molecular integrals, which include the one-electron integral $\mathbf{h}$, the two-electron integral $\mathbf{K}^{lk}$, the effective one-electron integrals $\mathbf{h}^{\text{eff}}$, $\mathbf{h}^{\text{eff}'}$, and $\bar{\mathbf{h}}^{\text{eff}}$ (see Ref. 27 the definitions of these terms), the Fock matrix with all the active orbitals fully occupied $\mathbf{f}^h$, and the generalized Fock matrix $\mathbf{f}$. Below, we present examples (the $S_{rs}^{(-2)}$ subspace) of the pseudensities and CI derivatives for both SC-NEVPT2 and PC-NEVPT2. Of course, the pseudodensities and CI derivatives for other subspaces can be derived in a similar manner.

**SC-NEVPT2 Source Term.** For SC-NEVPT2, the derivations can be further simplified by writing a Lagrangian,

$$L^{\text{SC-NEVPT2}} = E + \sum_{lk} P_l^{(k)} \left( f_l^{(k)} - N_l^{(k)} \right) + \sum_{lk} Q_l^{(k)} \left( g_l^{(k)} - \varepsilon_l^{(k)} \right) + \sum_{lk} R_l^{(k)} (h_l^{(k)} - \Delta_l^{(k)}), \tag{31}$$

where $\varepsilon_l^{(k)}$ is the denominator in the active space $\Delta_l^{(k)}$ is the difference between the virtual and core orbital energies; $f_l^{(k)}$, $g_l^{(k)}$, and $h_l^{(k)}$ are the explicit mathematical forms of $N_l^{(k)}$, $\varepsilon_l^{(k)}$, and $\Delta_l^{(k)}$, respectively; and $P_l^{(k)}$, $Q_l^{(k)}$, and $R_l^{(k)}$ are the corresponding multipliers. For example, in the $S_{rs}^{(-2)}$ subspace,

$$f_{rs}^{(-2)} = \frac{1}{2} \sum_{aba'b'} (a'r|b's)\Gamma_{ab,a'b'}^{(2)}(ar|bs), \tag{32}$$

$$g_{rs}^{(-2)} = \frac{1}{2} \sum_{aba'b'} (a'r|b's)K_{ab,a'b'}(ar|bs), \tag{33}$$

$$h_{rs}^{(-2)} = \varepsilon_r + \varepsilon_s. \tag{34}$$

where $\mathbf{\Gamma}^{(2)}$ is the two-electron RDM, $\mathbf{K}$ is the extended Koopmans' matrix for double ionization,[27,58]

$$K_{a'b',ab} = \sum_c \Gamma^{(2)}_{a'c,b'b} h^{\text{eff}}_{ac} + \sum_c \Gamma^{(2)}_{a'a,b'c} h^{\text{eff}}_{bc}$$
$$+ \frac{1}{2}\left[(ec|ad)\left(2\Gamma^{(3)}_{a'd,b'b,ec} + \delta_{be}\Gamma^{(2)}_{a'd,b'c}\right)\right] \quad (35)$$
$$+ \frac{1}{2}\left[(ec|bd)\left(2\Gamma^{(3)}_{a'a,b'd,ec} + \delta_{ae}\Gamma^{(2)}_{a'c,b'd}\right)\right],$$

and $\varepsilon_r$ is the eigenvalue of the generalized Fock operator for the orbital $r$; the negative sign has been dropped for simplicity. These multipliers are then

$$P^{(-2)}_{rs} = 2T^{(-2)}_{rs} + T^{(-2)}_{rs}\Delta^{(-2)}_{rs}, \quad (36)$$

$$Q^{(-2)}_{rs} = -T^{(-2)2}_{rs}, \quad (37)$$

$$R^{(-2)}_{rs} = N^{(-2)}_{rs} T^{(-2)2}_{rs}. \quad (38)$$

Then, the pseudodensity out of the active space is simply

$$d^{\text{Fock}}_{rr} = d^{\text{Fock}}_{ss} = R^{(-2)}_{rs}. \quad (39)$$

These definitions of the multipliers allow the number of mathematical operations to be minimized by separating the active space operations from the rest of the operations. For instance, the second and third terms in Eq. (31) for the $S^{(-2)}_{rs}$ subspace can be rewritten as

$$\sum_{rs} P^{(-2)}_{rs}\left[f^{(-2)}_{rs} - N^{(-2)}_{rs}\right] + \sum_{rs} Q_{rs}\left[g^{(-2)}_{rs} - \varepsilon^{(-2)}_{rs}\right]$$
$$= \sum_{a'a,b'b} \Gamma^{(2)}_{a'a,b'b} \sum_{rs}(a'r|b's)(ar|bs)P^{(-2)}_{rs} - \sum_{rs} P^{(-2)}_{rs}N^{(-2)}_{rs} \quad (40)$$
$$+ \sum_{a'a,b'b} K_{a'b',ab} \sum_{rs}(a'r|b's)(ar|bs)Q^{(-2)}_{rs} - \sum_{rs} Q^{(-2)}_{rs}\varepsilon^{(-2)}_{rs},$$

and the two-particle pseudodensity is

$$D_{ar,bs} = P^{(-2)}_{rs} \sum_{a'b'}(a'r|b's)\Gamma^{(2)}_{ab,a'b'} + Q^{(-2)}_{rs} \sum_{a'b'}(a'r|b's)K_{ab,a'b'}. \quad (41)$$

Additionally, by forming the intermediate matrices

$$M^{(-2)}_{a'a,b'b} = \sum_{rs}(a'r|b's)(ar|bs)P^{(-2)}_{rs}, \quad (42)$$

$$E^{(-2)}_{a'a,b'b} = \sum_{rs}(a'r|b's)(ar|bs)Q^{(-2)}_{rs}, \tag{43}$$

the pseudodensity in the active space is simply

$$d^{\text{eff}}_{ac} = \sum_{a'b'b} E_{a'a,b'b}\Gamma^{(2)}_{a'c,b'b}, \tag{44}$$

$$d^{\text{eff}}_{bc} = \sum_{a'ab} E_{a'a,b'b}\Gamma^{(2)}_{a'a,b'c}, \tag{45}$$

$$D_{ec,ad} = \frac{1}{2}\sum_{a'b'b} E_{a'a,b'b}\left(2\Gamma^{(3)}_{a'd,b'b,ec} + \delta_{be}\Gamma^{(2)}_{a'd,b'c}\right), \tag{46}$$

$$D_{ec,bd} = \frac{1}{2}\sum_{aa'b'} E_{a'a,b'b}\left(2\Gamma^{(3)}_{a'a,b'd,ec} + \delta_{ae}\Gamma^{(2)}_{a'c,b'd}\right), \tag{47}$$

and the CI derivative associated with this term is

$$\begin{aligned} y_I &= \sum_{a'a,b'b} M_{a'a,b'b}\Gamma^{(2)I}_{a'a,b'b} + \sum_{a'a,b'b} E_{a'a,b'b}\times\left\{h^{\text{eff}}_{ac}\Gamma^{(2)I}_{a'c,b'b} + h^{\text{eff}}_{bc}\Gamma^{(2)I}_{a'a,b'c}\right. \\ &\quad + \frac{1}{2}\left[(ec|ad)\left(2\Gamma^{(3)I}_{a'a,b'b,ec} + \delta_{be}\Gamma^{(2)I}_{a'd,b'c}\right)\right] \\ &\quad + \left.\frac{1}{2}\left[(ec|bd)\left(2\Gamma^{(3)I}_{a'a,b'd,ec} + \delta_{ae}\Gamma^{(2)I}_{a'c,b'd}\right)\right]. \end{aligned} \tag{48}$$

where $\Gamma^I$ is the derivative of the RDM with respect to the CI coefficient (RDM derivative). For example,

$$\Gamma^{(2)I}_{a'a,b'b} = \langle I|\hat{E}_{a'a,b'b}|\Psi^{(0)}\rangle + \langle\Psi^{(0)}|\hat{E}_{a'a,b'b}|I\rangle. \tag{49}$$

Thus, for a moderately sized active space, the most difficult step in forming the density is the formation of the intermediate matrix for $V^{(-1)}_{rsi}$ subspace, which scales with $O(N^2_{\text{vir}}N_{\text{clo}}N^2_{\text{act}})$. When the size of the active space is large, the intermediate matrix formation step for $V'^{(-1)}_r$, which scales with $O(N_{\text{vir}}N^6_{\text{act}})$, and the corresponding derivative evaluation, $O(N^9_{\text{act}})$, is dominant term. RDM derivatives up to the fourth order are needed, and generating these derivatives is a major computational bottleneck in our current algorithm. The active space derivative evaluation steps are parallelized using the DGEMM function in BLAS (basic linear algebra subprograms).

**PC-NEVPT2 Source Term.** The PC-NEVPT2 energy expression is quite simple, and a separate Lagrangian is not necessary. For instance, the explicit expression for the energy in the $S_{rs}^{(-2)}$ subspace is

$$E_{rs}^{(-2)} = \sum_{rs}\sum_{ab}\sum_{\mu}\left[2T_{rs,\mu}(rb|sa)\sum_{a'b'}V_{\mu,a'b'}\Gamma_{a'b',ab}^{(2)} \right.$$
$$\left. +\sum_{\nu}T_{rs,\mu}V_{\mu,a'b'}\left\{K_{a'b',ab}+\Gamma_{a'b',ab}^{(2)}(\varepsilon_r+\varepsilon_s)\right\}V_{\nu,ab}T_{rs,\nu}\right]. \quad (50)$$

Then, when orthogonal basis functions are used, the pseudodensity out of the active space is

$$d_{rr}^{Fock} = d_{ss}^{Fock} = \sum_{\mu}T_{rs,\mu}^2. \quad (51)$$

The two-particle pseudodensity is then

$$D_{ar,bs} = 2\sum_{\mu}T_{rs,\mu}\sum_{a'b'}V_{\mu,a'b'}\Gamma_{a'b',ab}^{(2)}. \quad (52)$$

By forming the intermediate vector

$$r_{ab,rs} = \sum_{\mu}V_{\mu,ab}T_{rs,\mu}, \quad (53)$$

and the intermediate matrix

$$M_{a'a,b'b}^{(-2)} = \sum_{rs}2r_{a'b',rs}r_{ab,rs}(\varepsilon_r+\varepsilon_s)+4(a'r|b's)r_{ab,rs}, \quad (54)$$

$$E_{a'a,b'b}^{(-2)} = \sum_{rs}2r_{a'b',rs}r_{ab,rs}, \quad (55)$$

the CI derivative and the pseudodensities in the active space can be similarly evaluated as in the strongly contracted case. Therefore, the computational routine for SC-NEVPT2 can be applied here as well. These definitions of the intermediate vectors and matrices make the computational scaling for evaluating these quantities the same as those in the strongly contracted case; however, the computational cost itself is higher due to the increased number of perturbation functions. Finally, we note that we can reduce the number of operations by introducing restrictions on the closed and virtual summation indices as in the case of energy evaluation,[27] for $S_{ij,rs}^{(0)}$, $S_{rs,i}^{(-1)}$, $S_{ij,r}^{(1)}$, $S_{rs}^{(-2)}$, and $S_{ij}^{(2)}$ subspaces for both SC- and PC-NEVPT2.

**Finalizing Gradient Evaluation.** Before solving the Z-vector equation, the pseudocanonical condition [the last line in Eq. (23)] should be considered with the multipliers

$$z_{ij}^c = -\frac{1}{2}\frac{Y_{ij} - Y_{ji}}{\varepsilon_i - \varepsilon_j}, \tag{56}$$

$$z_{rs}^c = -\frac{1}{2}\frac{Y_{rs} - Y_{sr}}{\varepsilon_r - \varepsilon_s}. \tag{57}$$

After the Z-vector equation is solved, we use the obtained **Z** and **z** to obtain the relaxed densities by the method described in previous works.[37,38] The density-fitting scheme is used to evaluate the two-electron integrals and their derivatives.[38,59,60] Some separable terms in the two-electron integrals (due to the effective one-electron integrals) should be added to result of the final gradient.

## 4. NUMERICAL EXAMPLES

An in-house program for the NEVPT2 analytical gradient was interfaced with the open-source program BAGEL.[18] In all calculations, the cc-pVTZ basis set and its corresponding density-fitting JKFIT basis sets were used unless otherwise noted. For comparative purposes, the single-state CASPT2 and CASSCF geometries were also optimized. In the CASPT2 calculations, a real level shift[61] of 0.2 $E_h$ was applied to avoid intruder states. All the times reported here were measured using Intel Xeon Gold 6140 2.3 GHz processors (2 CPUs and 36 physical cores per node).

Using the SC- and PC-NEVPT2 gradient, we optimized the ground-state geometries of acrolein, benzene, benzyne (both singlet and triplet), PSB3 (model of the rhodopsin chromophore, penta-2,4-dieniminium cation), *p*HBI (model of the GFP chromophore, *para*-hydroxybenzilideneimidazolin-5-one anion), and porphine. The optimized ground-state geometries using PC-NEVPT2 are shown in Fig. 1, and the root-mean-squared (RMS) distances to the SC-NEVPT2, CASSCF and CASPT2 geometries are shown in Table 1. First, one can notice that SC-NEVPT2 geometry optimizations did *not* converge in the cases of benzene and porphine. This is due to the lack of invariance in SC-NEVPT2: The SC-NEVPT2 energy is not invariant with respect to the rotations between inactive orbitals.[28,52] We will discuss this effect later. On the other hand, PC-NEVPT2 optimizations converged smoothly to equilibrium geometries. The NEVPT2-optimized geometries are 2 to 5 times closer to the CASPT2 geometries than they are to the CASSCF geometries based on their RMS distances, except for porphine, where the cc-pVDZ basis set was used for the CASPT2 calculation; if the same cc-pVDZ basis set was used, the root-mean-squared (RMS) difference between PC-NEVPT2 and CASPT2 geometries is 0.001 Å. This result highlights the importance of the selection of an appropriate basis set to ensure the accuracy of the geometry optimization. We also computed the singlet–triplet splitting in benzyne with a geometry relaxation effect (3.36 kcal/mol), which is 0.30 kcal/mol smaller than the value previously reported without geometry relaxation.[62] The times required for the gradient evaluations for each geometry optimization step, except the reference CASSCF calculations, are shown in Table 2 along with the sizes of the active spaces and the number of basis functions.

To further investigate the effect of non-invariance in SC-NEVPT2, we compared the analytical gradients with the corresponding numerical gradient for lithium fluoride (LiF), acrolein, benzene, and 1,2,4-pyrazine using the SVP basis. The resulting errors are shown in Table 3. The PC-NEVPT2 analytical gradients and numerical gradients agree within an error of 5.0×10$^{-6}$ a.u. even when the numerical gradients are computed by a two-point finite difference. This situation is the same for SC-NEVPT2, except for the case of benzene. More specifically, the SC-NEVPT2 analytical gradients agree well with the numerical gradients for the system is small (LiF), or without obvious symmetry (acrolein, 1,2,4-pyrazine), but they do not agree when there are degenerate inactive orbitals (benzene). When we lift the symmetry by substituting the 1,2,4-carbon atoms with nitrogens (1,2,4-pyrazine) or moving them by 0.5 angstrom ("distorted benzene"), the analytical and numerical gradients in SC-NEVPT2 agree well. Because of the quality of the analytical gradient, the geometry optimization did not converge with SC-NEVPT2 analytical gradient when there is a symmetry. This result is consistent with the results of the local correlation implementation of NEVPT2,[52] and this seems to stem from the lack of the invariance in the SC-NEVPT2 energy.[28] Thus, we conclude that PC-NEVPT2 is more suitable for optimizing the molecular structures in large systems, in which they are prone to the accidental degeneracies between the orbitals.

We also parallelized our algorithm. Because the amplitudes do not need to be stored to solve the amplitude equations and evaluate the density matrices in NEVPT2, distributing the algorithm into the MPI processes is straightforward. The calculations are distributed by the virtual and occupied indices. As shown in Fig. 2, we have tested the parallel performance of our implementation. The calculations with 6 MPI processes (246.6 s) were approximately 3.3 times faster than those with a single MPI process (816.5 s). In particular, the NEVPT2 energy and density matrix evaluations were notably accelerated by parallelization [calculations with 6 MPI processes (93.1 s) were 3.9 times faster than those with a single MPI process (365.2 s)]. We note that the parallel performance of our program is better when the number of nodes equals the number of MPI processes (when using up to 54 CPU cores). Even for the largest system tested here (916 basis functions), the memory consumption was small enough that the calculation could fit in a single node with 192 GB of memory.

## 5. CONCLUSIONS AND FUTURE PROSPECTS

In this work, the analytical gradient theory for single-state NEVPT2 was implemented for both the strongly contracted (SC) and partially contracted (PC) variants. Unlike PC-NEVPT2, the analytical gradient theory for SC-NEVPT2 was numerically unstable especially when inactive orbitals are degenerate. We performed geometry optimizations using NEVPT2, and the resulting molecular structures were comparable to those obtained by CASPT2. The algorithm was optimized by separating the active space operations from the other operations. This kind of separation can be applied for the gradient algorithms for other multireference theories. The algorithm was parallelized, and the process can be readily distributed into many processes.

There are many possible extensions of current algorithms, such as multistate perturbation theory [quasidegenerate NEVPT2 (QD-NEVPT2)] for studying excited-state manifolds.[31] Most likely, the machinery that is developed for (X)MS-CASPT2 would be similarly applicable. We also note that the current algorithm can be applied only to systems with small active spaces because the derivatives of the fourth-order RDM with respect to the CI coefficients must be determined, and this is given by

$$\Gamma^{(4)I}_{a'a,b'b,c'c,d'd} = \langle I | \hat{E}_{a'a,b'b,c'c,d'd} | \Psi^{(0)} \rangle + \langle \Psi^{(0)} | \hat{E}_{a'a,b'b,c'c,d'd} | I \rangle. \tag{58}$$

For the (6$e$, 6$o$) active space, this requires 5 GB of memory (double precision), but this quickly becomes intractable with larger active spaces [658 GB for (8$e$, 8$o$)]. This situation does not improve significantly even if symmetry, hard-disk interface, or multipassing algorithms are exploited. Nevertheless, the NEVPT2 method offers clear advantages, such as separability and avoiding the intruder state problem. Thus, there are many opportunities for improving the current algorithm, and further improvements will be reported in a due course.

## ACKNOWLEDGMENTS

We thank Prof. Toru Shiozaki (Northwestern) for insightful discussions. The debugging of the PC-NEVPT2 energy algorithm was facilitated by the existing implementation of the SC-NEVPT2 algorithm in BAGEL.[18,57] This work was supported by the National Research Foundation of Korea (NRF) grant funded by the Korean government (MSIT) (No. 2019R1C1C1003657).

K.; Ligabue, A.; Lutnæs, O. B.; Melo, J. I.; Mikkelsen, K. V.; Myhre, R. H.; Neiss, C.; Nielsen, C. B.; Norman, P.; Olsen, J.; Olsen, J. M. H.; Osted, A.; Packer, M. J.; Pawlowski, F.; Pedersen, T. B.; Provasi, P. F.; Reine, S.; Rinkevicius, Z.; Ruden, T. A.; Ruud, K.; Rybkin, V. V.; Sałek, P.; Samson, C. C. M.; de Merás, A. S.; Saue, T.; Sauer, S. P. A.; Schimmelpfennig, B.; Sneskov, K.; Steindal, A. H.; Sylvester-Hvid, K. O.; Taylor, P. R.; Teale, A. M.; Tellgren, E. I.; Tew, D. P.; Thorvaldsen, A. J.; Thøgersen, L.; Vahtras, O.; Watson, M. A.; Wilson, D. J. D.; Ziolkowski, M.; Ågren, H., The Dalton quantum chemistry program system. *WIREs Comput. Mol. Sci.* **2014,** *4*, 269-284.

36. Dyall, K. G., The choice of a zeroth-order Hamiltonian for second-order perturbation theory with a complete active space self-consistent-field reference function. *J. Chem. Phys.* **1995,** *102*, 4909-4918.

37. Celani, P.; Werner, H.-J., Analytical Energy Gradients for Internally Contracted Second-Order Multireference Perturbation Theory. *J. Chem. Phys.* **2003,** *119*, 5044-5057.

38. Györffy, W.; Shiozaki, T.; Knizia, G.; Werner, H.-J., Analytical Energy Gradients for Second-Order Multireference Perturbation Theory using Density Fitting. *J. Chem. Phys.* **2013,** *138*, 104104.

39. MacLeod, M. K.; Shiozaki, T., Communication: Automatic Code Generation Enables Nuclear Gradient Computations for Fully Internally Contracted Multireference Theory. *J. Chem. Phys.* **2015,** *142*, 051103.

40. Park, J. W.; Shiozaki, T., Analytical Derivative Coupling for Multistate CASPT2 Theory. *J. Chem. Theory Comput.* **2017,** *13*, 2561-2570.

41. Park, J. W.; Shiozaki, T., On-the-Fly CASPT2 Surface-Hopping Dynamics. *J. Chem. Theory Comput.* **2017,** *13*, 3676-3683.24

**FIGURES**

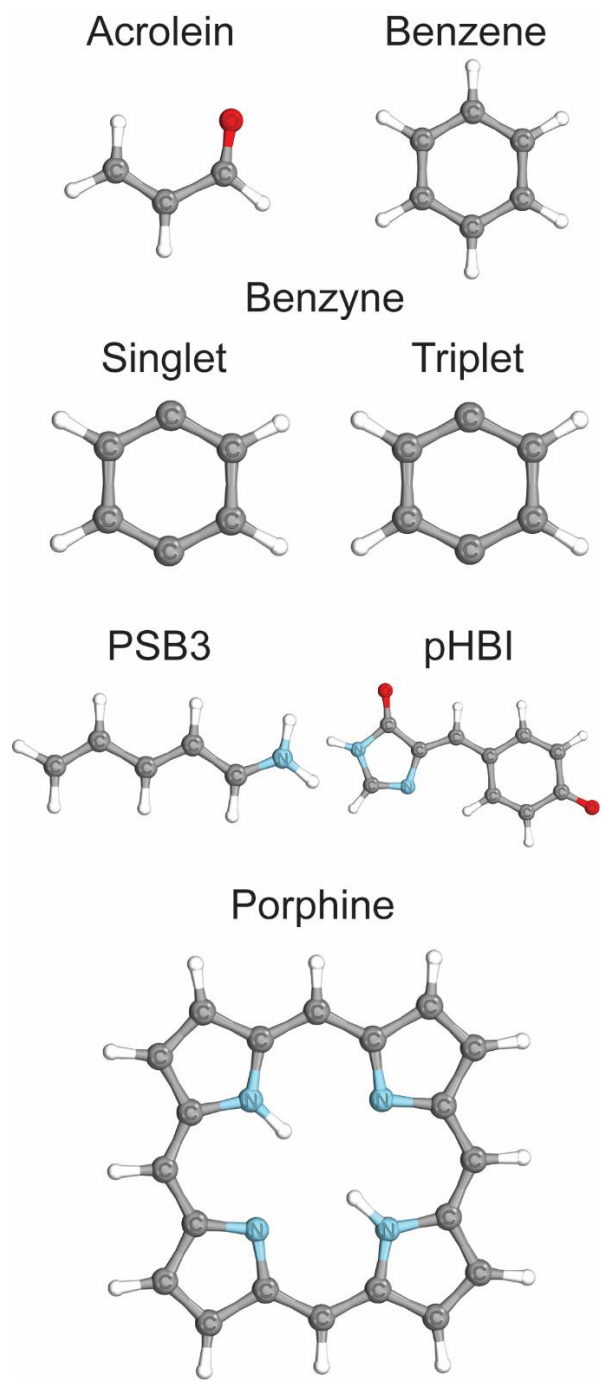

**FIGURE 1.** The PC-NEVPT2-optimized ground-state geometries of acrolein, benzene, benzyne (singlet and triplet), PSB3, *p*HBI, and porphine. See the Supporting Information (SI) for their Cartesian coordinates. The molecular graphics are created using the software IboView.[63,64]

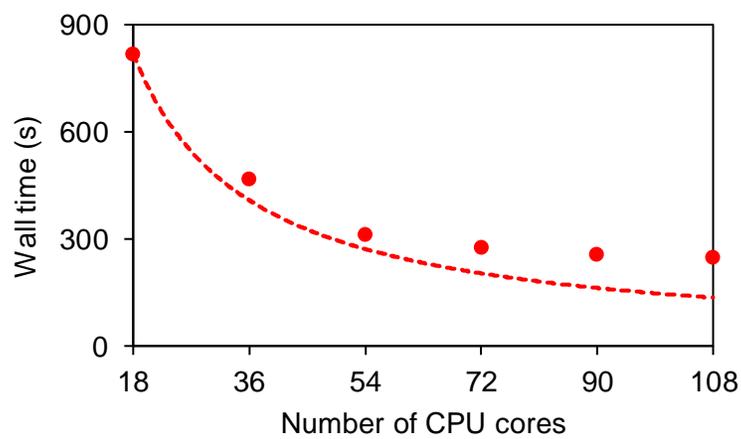

**FIGURE 2.** The parallel performance of the PC-NEVPT2 gradient of porphine [active space: (4*e*,4*o*)] (strong scaling; the dashed lines are the ideal scaling). Eighteen physical cores were used per MPI process.

# TABLES

**TABLE 1.** The root-mean-squared (RMS) distances (Å) from the PC-NEVPT2 geometries to the SC-NEVPT2, CASSCF and CASPT2 geometries. The cc-pVTZ basis set was used. The Cartesian coordinates of the optimized geometries are listed in the Supporting Information.

|          | active space | SC-NEVPT2 | CASSCF | CASPT2 |
|----------|--------------|-----------|--------|--------|
| Acrolein | (6$e$,5$o$)  | 0.005     | 0.027  | 0.006  |
| Benzene  | (6$e$,6$o$)  | [a]       | 0.005  | 0.002  |
| PSB3     | (6$e$,6$o$)  | 0.004     | 0.033  | 0.007  |
| $p$HBI   | (4$e$,3$o$)  | 0.001     | 0.045  | 0.009  |
| Porphine | (4$e$,4$o$)  | [a]       | 0.028  | 0.032[b] |

[a] Did not converge, due to the numerical instability.
[b] The cc-pVDZ basis set was used for CASPT2 calculations due to the restrictions in the computational resources available in our group. The RMS distance between CASPT2 cc-pVDZ porphine and PC-NEVPT2 cc-pVDZ porphine is 0.001 Å.

**TABLE 2.** The sizes of the systems and the timing for the PC-NEVPT2 gradient using 36 CPU cores. The cc-pVTZ basis set and corresponding JKFIT basis were used.

|          | Active space | $N_{bas}^a$ | $N_{aux}^b$ | Wall time[c] |
|----------|--------------|-------------|-------------|--------------|
| Acrolein | (6e,5o)      | 176         | 436         | 5.63         |
| Benzyne  | (2e,2o)      | 236         | 594         | 13.1         |
| Benzene  | (6e,6o)      | 264         | 654         | 20.6         |
| PSB3     | (6e,6o)      | 292         | 714         | 30.76        |
| pHBI     | (4e,3o)      | 518         | 1316        | 77.96        |
| Porphine | (4e,4o)      | 916         | 2316        | 467          |

[a] Number of basis functions.
[b] Number of auxiliary basis functions for density-fitting (DF) approximations.
[c] Times include the correlation energy and gradient evaluations, including Z-vector solution. This time does not include the time required for CASSCF reference calculations.

**TABLE 3.** The root-mean-squared (RMS) errors between the analytical and numerical gradients. The SVP basis set and corresponding JKFIT basis were used. The numerical gradients were evaluated using a two-point finite difference formula with a finite difference (d$x$) of 0.001 bohr. The Cartesian coordinates of the tested geometries, as well as evaluated gradients, are listed in the Supporting Information.

| System | Active space | SC-NEVPT2 | PC-NEVPT2 |
|---|---|---|---|
| LiF | (4$e$,6$o$) | 2.39×10$^{-6}$ | 3.01×10$^{-6}$ |
| Acrolein | (6$e$,5$o$) | 3.44×10$^{-6}$ | 3.36×10$^{-6}$ |
| 1,2,4-Pyrazine | (6$e$,6$o$) | 1.28×10$^{-6}$ | 1.29×10$^{-6}$ |
| Benzene | (6$e$,6$o$) | 1.69×10$^{-3}$ | 7.96×10$^{-7}$ |
| Distorted Benzene | (6$e$,6$o$) | 1.00×10$^{-5}$ | 7.89×10$^{-7}$ |